\documentclass[preprint]{elsarticle}
\usepackage{multicol}
\usepackage{tikz}
\usepackage{empheq}
\usepackage{algpseudocode}
\newcounter{casenum}

\usetikzlibrary{patterns,positioning}
\usetikzlibrary{shapes.geometric,calc,decorations.text}
\usepackage{caption}
\usepackage{subcaption}
\usepackage{todonotes}
\usepackage{bbold} 
\usepackage{float}
\usepackage{amsmath}
\usepackage{amsfonts,amssymb,amsmath,amsthm}
\usepackage{xcolor}
 \usepackage{url}

\usepackage[justification=centering]{caption}
\usetikzlibrary{patterns,positioning}

\usepackage{caption}

\usepackage{amsmath}
\usepackage{amsfonts,amssymb,amsmath,amsthm}
\usepackage{anysize} 

\usetikzlibrary{shapes.misc}

\usetikzlibrary{shapes.geometric}
\usetikzlibrary{patterns,positioning}
\usepackage{amsmath, amssymb}
\usepackage{tikz}
\usetikzlibrary{shapes.geometric,calc,decorations.text}
\usetikzlibrary{patterns}

\usepackage{mathrsfs}
\usepackage{amssymb}
\usepackage{dsfont}
\usetikzlibrary{patterns}
\usetikzlibrary{shapes.geometric}
\def\mytransformation{
\pgfmathsetmacro{\myX}{0.9*\pgf@x}
\pgfmathsetmacro{\myY}{0.0045*(\pgf@x)*(\pgf@x)+(\pgf@y)}
\setlength{\pgf@x}{\myX pt}
\setlength{\pgf@y}{\myY pt}
}

\makeatother
\usetikzlibrary{backgrounds}
\usetikzlibrary{shapes.geometric}
\usetikzlibrary{calc}
\usepackage{enumerate}
\usepackage{mathrsfs}
\usepackage{amssymb}
\usepackage{graphicx}%
\usepackage{multirow}%
\usepackage{amsmath,amssymb,amsfonts}%
\usepackage{amsthm}%
\usepackage{mathrsfs}%
\usepackage[title]{appendix}%
\usepackage{xcolor}%
\usepackage{textcomp}%
\usepackage{manyfoot}%
\usepackage{algorithm}%
\usepackage{algorithmicx}%
\usepackage{algpseudocode}%
\usepackage{setspace}
\usepackage{pdflscape}
\usepackage{graphicx}
\usepackage{pst-node}
\usepackage{booktabs}
\usepackage{tikz}
\usepackage{todonotes}

\renewenvironment{proof}{{\it Proof.}}{\qed \medskip}
\newtheorem{theorem}{Theorem}[section]

\newtheorem{proposition}[theorem]{Proposition}
\newtheorem{obs}[theorem]{Observation}
\newtheorem{remark}[theorem]{Remark}
\newtheorem{claim}[theorem]{Claim}

\newtheorem{lemma}[theorem]{Lemma}
\newtheorem{definition}[theorem]{Definition}
\newtheorem{corollary}[theorem]{Corollary}

\usepackage{mathtools}

\newcommand{\cone}{\textup{cone}}

\newcommand{\proj}{\textup{proj}}

\newcommand{\textoverline}[1]{$\overline{\mbox{#1}}$}

\begin{document}

\title{Extended formulations for the maximum weighted co-2-plex problem}
\author[1,2]{Alexandre Dupont-Bouillard\corref{cor1}}
\ead{dupont-bouillard@lipn.fr}
\author[2]{Pierre Fouilhoux}
\ead{fouilhoux@lipn.fr}
\author[3]{Roland Grappe}
\ead{roland.grappe@lamsade.dauphine.fr}
\author[2]{Mathieu Lacroix}
\ead{lacroix@lipn.fr}

\affiliation[1]{\organization={LIMOS CNRS UMR 6158 Université Clermont Auvergne},   \adressline={1 Rue de la Chebarde}, \city={Aubière}, \postcode={63178}, \country={France}}

\affiliation[2]{\organization={LIPN CNRS UMR 7030 Université Sorbonne Paris Nord},  \adressline={99 Av. Jean Baptiste Clément}, \city={Villetaneuse}, \postcode={93430}, \country={France}}

\affiliation[3]{\organization={LAMSADE CNRS UMR 7243 Université Paris Dauphine},   \adressline={Pl. du Maréchal de Lattre de Tassigny}, \city={Paris}, \postcode={75016}, \country={France}}
\cortext[cor1]{Corresponding author}

\begin{abstract}
Given an input graph and weights on its vertices, the maximum co-2-plex problem is to find a subset of vertices maximizing the sum of their weights and inducing a graph of degree at most 1. In this article, we analyze polyhedral aspects of the maximum co-2-plex problem.
The co-2-plexes of a graph are known to be in bijection with the stable sets of an auxiliary graph called the utter graph~\cite{dupontbouillard2024contractions}. 
We use this bijection to characterize contraction perfect graphs' co-2-plex polytopes in an extended space. It turns out that the total dual integrality of the associated linear system also characterizes contraction perfectness of the input graph.

By projecting this extended space formulation, we obtain the natural variable space description of the co-2-plex polytopes of trees.
This projection yields a new family of valid inequalities for the co-2-plex polytope and we characterize when they define facets. 
Moreover, we show that these inequalities can be separated in polynomial time.
We characterize the graphs for which this formulation describes an integer polytope.

These linear systems are extended to valid integer linear programs (ILPs) for the maximum co-2-plex problem whose linear relaxation values are tighter than the state of the art for this problem~\cite{bala}. Finally, we provide an experimental comparison of several implementations of our new ILP formulations with the state-of-the-art ILP for this problem and analyze their respective performances relatively to the density of the input graphs.
\end{abstract}

\maketitle

\section*{Introduction}
All the graphs in this paper are simple and connected. 
Given a graph $G=(V,E)$, we denote its \textit{complement} by $\overline{G}=(V,\overline E)$, where $\overline{E} = \{ e \in \binom{V}{2} : e \notin E \}$. 
We respectively denote by $V(G)$ and $E(G)$ the vertex set and the edge set of $G$. Two vertices $u$ and $v$ are \emph{adjacent} if $uv \in E(G)$. 
Given a subset of vertices $W \subseteq V$, let $E(W)$ denote the set of edges of $G$ having both endpoints in $W$ and $\delta (W)$ the set of all edges having exactly one endpoint in $W$.
 When $W$ is a singleton $\{w\}$, we will simply write $\delta (w)$. We say that the edges in $\delta (w)$ are \textit{incident} to $w$, and two edges sharing an extremity are said {\em adjacent}.
For $F \subseteq E$, let $V(F)$ denote the set of vertices incident to any edge of $F$.
Given $W\subseteq V$, the graph $G[W]=(W,E(W))$ is the \textit{subgraph of $G$ induced by $W$}. When $H$ is an induced subgraph of $G$, we say that $G$ {\em contains} $H$. 
Given a vertex $u\in V(G)$, we denote  by $N_G(u) = \{w \in V(G) : uw \in E(G) \}$ its \textit{neighborhood}, and by $N_G[u] = N_G(u) \cup \{ u\}$ its \textit{closed neighborhood}, when clear from context we simply write $N(u)$ and $N[u]$. Two vertices $u$ and $v$ are {\em true twins} if $N[u] =N[v]$ and {\em false twins} if $N(u) = N(v)$ and $uv \notin E$.
A vertex $w$ is \emph{complete} to a subset of vertices $W$ if it is adjacent to each vertex of $W$ and does not belong to $W$. A vertex $w$ is \emph{anticomplete} to $W$ if it is complete to $W$ in the complement graph.
 A vertex $u$ complete to $V \setminus \{u\}$ is said to be \emph{universal}.
 
A subset $K$ of vertices is a \emph{$k$-plex} (resp. \emph{co-$k$-plex}) if $G[K]$ has minimum degree $|K| - k$ (resp. maximum degree $k-1$). Note that $k$-plexes are complements of co-$k$-plexes. \emph{Cliques} are $1$-plexes and \emph{stable sets} are co-1-plexes. Note that a co-2-plex $S$ induces a set of isolated vertices and edges, for that reason, we consider the \emph{vertex edge representation} $(W,F)$ of $S$, so that $W \subseteq S$ is its set of isolated vertices and $F = E(S)$ its set of isolated edges.
The size of the largest clique (resp. stable set) is denoted by $\omega (G)$ (resp. $\alpha (G)$). We also denote by $\alpha_2(G)$ the size of the largest co-2-plex of $G$. A \emph{coloring} of a graph is an assignment of colors to each vertex such that every pair of adjacent vertices have different colors. We denote by $\chi (G)$ the minimum number of colors of a coloring of $G$.

A \textit{path} (resp. \textit{hole}) is a graph induced by a sequence of vertices $(v_1,\dots,v_p)$ whose edge set is $\{ v_iv_{i+1}:  \ i = 1,\dots, p-1 \} $ (resp. $\{ v_iv_{i+1}:  \ i = 1,\dots, p-1 \} \cup \{v_1v_p \}$ with $p\ge4$). Note that this definition usually corresponds to induced paths. 
A subset of vertices induces a path (resp. hole) if its elements can be ordered into a sequence inducing a path (resp. hole).
An \emph{antipath} (resp. \textit{antihole}) of $G$ is a path (resp. hole) of $\overline{G}$. The {\em length} of a path or hole is its number of edges, and the length of an antipath or antihole is the length of its complement. A path, antipath, hole of antihole is called \emph{long} if its length is at least 5. The { \em parity} of a path, hole, antipath, or antihole is the parity of its length.
A graph is connected if there exists a path between each pair of vertices. A set of vertices $V$ (resp. edges $E$) is connected if $G[V]$ (resp. $(V(F),F)$) is.

The \textit{contraction} of an edge $uv\in E(G)$ yields a new graph $G/uv$ built from $G$ by deleting $u$, $v$, adding a new vertex $w$, and adding the edges $wz$ for all $z\in (N_G(u) \cup N_G(v)) \setminus \{ u,v\}$. 
For $F \subseteq E$, we denote by $G/F$ the graph obtained from $G$ by contracting all the edges in $F$. 
An edge $uv$ and a vertex $w$ are \textit{adjacent by contraction} if $w$ is adjacent to $x_{uv}$ in $G/uv$, where $x_{uv}$ is the vertex of $G/uv$ resulting from the contraction of $uv$.
In other words, at least one of $uw$ and $vw$ is in $E$.
Two non-adjacent edges $uv$ and $xy$ are \textit{adjacent by contraction} if contracting both edges results in two adjacent vertices, that is, $\delta(\{u,v\}) \cap \delta (\{x,y \})\neq \emptyset$. 

A graph $G$ is \emph{perfect} if $\chi (G') = \omega (G')$ holds for every induced subgraph $G'$ of $G$.
Perfect graphs are the graphs containing neither odd holes nor odd antiholes; this tremendous result is known as the strong perfect graph theorem~\cite{perfectgraphtheorem}. A graph is \emph{contraction perfect} if it is perfect and remains perfect under the 
contraction of any edge set~\cite{dupontbouillard2024contractions}.
A graph is chordal if it contains no hole. A \emph{tree} is a connected graph containing neither holes nor \emph{triangles}, that is, cliques of size three. Equivalently, trees are connected chordal graphs with no triangle. Note that chordal graphs are contraction perfect~\cite{dupontbouillard2024contractions}.
The converse does not hold as shows the hole of size 4.
\medskip

Let $\mathcal{P} = \{x \in \mathds{R}^n:Ax \le b\}\subseteq \mathds{R}^n$ be a \emph{polyhedron}, that is, the intersection of finitely many half-spaces. The \emph{dimension} of $\mathcal{P}\subseteq \mathds{R}^n$ denoted by dim $(\mathcal{P})$, is the maximum number of affinely independent points in $\mathcal{P}$ minus one. 
If $a \in \mathds{R}^n\setminus \{0\}$, $\alpha \in \mathds{R}$, then the inequality $a^\top x \le \alpha$ is said to be \emph{valid} for $\mathcal{P}$ if $\mathcal{P} \subseteq \{ x \in \mathds{R}^n : a^\top x \le \alpha \}.$ 
We say that a valid inequality $a^\top x \le \alpha$ \emph{defines a facet} of $\mathcal{P}$ if $\textrm{dim}(\mathcal{P} \cap \{x \in \mathds{R}^n : a^\top x = \alpha \})$ is equal to $\textrm{dim}~(\mathcal{P})$ minus one. 
The polyhedron $\mathcal{P} \subseteq \mathds{R}^n$ is \emph{full dimensional} when dim~$(\mathcal{P})$ is equal to $n$.
In that case, a linear system $Ax \le b$ that defines $\mathcal{P}$ is \emph{minimal} if each inequality of the system defines a facet of $\mathcal{P}$. Moreover, these \emph{facet-defining inequalities} are unique up to scalar multiplication. A \emph{face} of $\mathcal{P}$ is a polyhedron obtained by setting to equality some inequalities in $Ax\leq b$.
Equivalently, a face is the intersection of the supporting hyperplane of a valid inequality and $\mathcal{P}$.
\emph{Extreme points} of a polyhedron are its faces of dimension zero.
Each extreme point satisfies $\textrm{dim}~(\mathcal{P})$ linearly independent valid inequalities for $\mathcal{P}$ with equality.
A polyhedron defined by a linear system with right-hand side $\mathbb{0}$ is a \emph{cone}. A set of \emph{generators} $\mathcal{G}$ of a cone $\mathcal{C}$ is a set of points of $\mathcal{C}$ such that each element of $\mathcal{C}$ can be expressed as a nonnegative combination of $\mathcal{G}\text{'s}$ elements. A set of generators of finite size always exists.
A polyhedron $\mathcal{P} \subseteq \mathds{R}^n$ is said to be \emph{integer} if all its non-empty faces contain an integer point. 
The orthogonal \emph{projection} of a polyhedron $\mathcal{P} = \{ (x,y) \in \mathds{R}^{n\times m} : Ax + By \le c \}$ onto the $x$ variable space is defined by $\text{proj}_x(\mathcal{P}) = \{x \in \mathds{R}^n : \exists y \in \mathds{R}^m \ Ax+By \le c \} $.   
A \emph{polytope} is a bounded polyhedron, equivalently, it can be defined as the convex hull of a finite number of points $V$ and is denoted by $\textrm{conv} (V)$.

Given a 0/1 matrix $M\neq 0 $ of size $m\times n$, we say that $\textrm{conv} \{x \in \{0,1 \}^n : Mx \le \mathbb{1}\}$ is a \emph{packing} polytope. Moreover, $M$ is said to be \emph{perfect} if  $\{ x\in \mathds{R}^n : Mx \le \mathbb{1}, \ x \ge \mathbb{0}\}$ is integer.
  \medskip

A linear system is \emph{total dual integral} (TDI) if for any integer $w$ such that $\max \{ w^\top x : Ax \le b \}$ is finite, the following dual program has an optimal integer solution: 
$$\min \{ b^\top \pi :  A^\top \pi = w, \pi\geq \mathbb{0} \}. $$

\medskip 
Given a subset $W$ of vertices of $G=(V,E)$, we denote by $\chi^W$ the \emph{incidence vector} of $W$ as the element of $\{0,1\}^{V}$ such that $\chi^W_v=1$ if $v \in W$ and $\chi^W_v=0$ otherwise. 
Similarly, we denote by $\zeta^F$ the incidence vector of $F \subseteq E$ by the element of $\{ 0,1\}^{E}$ having a 1 in component $e$ if $e \in F$ and 0 otherwise. 
The \emph{stable set polytope} (resp. \emph{co-2-plex polytope}) of a graph is defined as the convex hull of its stable sets (resp. co-2-plexes) incidence vectors.


\subsection*{Motivations.}

To find cohesive subgroups in social networks, it is natural to search for the maximum clique. However, asking that all pairs of vertices are adjacent in such a subgroup may be too restrictive, and it could be sufficient that "most" of the elements are pairwise adjacent. For this reason, several notions of "relaxed cliques" have been defined, and $k$-plexes are one of these; see~\cite{PATTILLO20139} for a survey. Finding maximum $k$-plexes is NP-hard for any positive fixed $k$~\cite{bala} (this also holds for the maximum co-$k$-plex problem by complementing the graph).
This topic has been especially lively over the last few years: there are algorithms based on local search~\cite{DBLP:journals/heuristics/Pullan21,Chen2019CombiningRL,8794502,Jin2019EffectiveRL,Chen_Wan_Cai_Li_Chen_2020}, heuristics~\cite{doi:10.1142/S0218213019500155}, Branch-and-Bound algorithms, some of which incorporating machine learning ingredients~\cite{bala,10.1007/978-3-319-11812-3_21,DBLP:journals/corr/abs-2301-07300,DBLP:journals/dase/WangJYL17,10.14778/3565816.3565817,DBLP:journals/corr/abs-2208-05763}, quadratic models~\cite{Stetsyuk} and other exact algorithms~\cite{Zhou_Hu_Xiao_Fu_2021}. Additionally, a framework for covering a graph with a minimum number of relaxed cliques has been proposed in~\cite{roberto2}, its application to cover a graph with 2-plexes highly relies on an efficient algorithm for finding maximum weighted 2-plexes. On the polyhedral side, new facets and valid inequalities have been found~\cite{bala}, but the facets of the polytopes are only characterized for paths, holes, co-2-plexes, and 2-plexes~\cite{pol}.  

\subsection*{Contributions}

In this paper, we propose a polyhedral study of the co-2-plex polytope in an extended space based on facial properties of the stable set polytope of an auxiliary graph called the utter graph~\cite{dupontbouillard2024contractions}. 
This study goes through a new characterization of contraction perfect graphs based on the integrality and total dual integrality of a linear system of inequalities. Following a similar scheme, we also provide a new characterization of chordal graphs. By projecting these systems (both coincide on chordal graphs), we get rid of additional variables and obtain a linear system describing the co-2-plex polytope of trees. We prove that this linear system describes an integer polyhedron if and only if the graph is a tree or a hole of size multiple of 3. It turns out that this linear system yields a new family of valid inequalities for any graph, and we characterize when they define facets of the co-2-plex polytope.

Based on these results, we propose new ILPs for the maximum co-2-plex problem and show that their linear relaxation values are tighter than the state of the art's compact one~\cite{bala}; some are even tighter than the state of the art reinforced by 2-plex inequalities.
Finally, we compare several implementations, including Branch-and-Cuts, that take advantage of the facial properties highlighted and show that they challenge the state of the art on some instance sets in terms of running time and overtake it in its number of branching nodes and root node gap. This experimental comparison validates the applicability of our preliminary polyhedral study of the co-2-plex polytope, which may be of independent theoretical interest.

\section{Preliminaries}
In this section, we recall the literature on contraction perfect graphs, polyhedral results on stable sets and co-2-plex polytopes, ILP for finding maximum stable sets and co-2-plexes, and TDI systems.

We denote the co-2-plex polytope of a graph $G$ by $\mathcal{P}(G)$. The variables of the linear systems we consider will be associated with vertices or edges of the graph. 
In particular, the variables $x$ associated with the vertices will form the \emph{natural variable space}.

\subsection{Contraction perfect graphs and utter graphs}

The starting point of our work is the remark from~\cite{dupontbouillard2024contractions} that co-2-plexes of a graph are in bijection with the stable sets of an auxiliary graph called \emph{utter graph}. 

\begin{definition}[\cite{dupontbouillard2024contractions}]
The \emph{utter graph} $u(G)$ of a graph $G=(V,E)$ is the graph having $V\cup E$ as vertex set and for which two vertices are adjacent if their corresponding elements in $G$ are either incident, adjacent, or adjacent by contraction.    
\end{definition}

Remark that, given the vertex edge representation $(W,F)$ of a co-2-plex $S$ of $G$, $W\cup F$ is a stable set of $u(G)$. This yields the following.

\begin{obs}[\cite{dupontbouillard2024contractions}]\label{lem:bij}
	The co-2-plexes of $G$ are in bijection with the stable sets of $u(G)$.
\end{obs}

To make use of the facial results on perfect graphs' stable set polytopes, one should ask when the utter graph is perfect, which has already been answered as stated in the next theorem.

\begin{theorem}[\cite{dupontbouillard2024contractions}]\label{the:matching}
A graph is contraction perfect if and only if its utter graph is perfect. 
\end{theorem}

\begin{theorem}[\cite{dupontbouillard2024contractions}]\label{the:contractsingleedges}
A perfect graph is contraction perfect if and only if it remains perfect under the contraction of any single edge.
\end{theorem}

\subsection{Polyhedra and linear systems}

First, as the empty set and all singletons are both stable sets and co-2-plexes, the following holds.

\begin{remark}\label{rem:dimension}
    Stable set polytopes and co-2-plex polytopes are full-dimensional.
\end{remark}

A row $m$ of a 0/1 matrix $M$ is said to be \emph{redundant} if $m \le m'$ holds component-wise for some row $m'$ of $M$.
We recall that perfect graphs are in bijection with perfect matrices up to row and column permutations.

\begin{theorem}[\cite{CHVATAL1975138},\cite{LOVASZ1972253}] \label{the:chva}
For a $0/1$ matrix $A$ with no redundant rows of size $n\times m$ with no column of $0$'s, the following statements are equivalent:
\begin{itemize}
    \item $A$ is the maximal clique-vertex incidence matrix of a perfect graph,
    \item $  \max \{ w^\top x : Ax \le \mathbb{1}, \ x \ge \mathbb{0} \} $ has an integral optimal solution $x$ for all $w \in \{0,1 \}^n$,
		\item The linear system $Ax \le \mathbb{1},$ $x\ge \mathbb{0}$ is TDI. 
    \item $A$ is perfect.
\end{itemize}

\end{theorem}

The third statement of Theorem~\ref{the:chva} can be written as follows.

\begin{corollary}\label{cor:perfect}
	A graph $G = (V,E)$ is perfect if and only if the following system is TDI:
	\begin{align}
		x(K) \le 1 && \text{for all maximal  cliques } K \text{ of } G, \label{eq:cliqueStableSetPolytope}\\
		-x_v \le 0 && \text{ for all } v \in V. \label{eq:trivialPositivity}
	\end{align}
\end{corollary}
TDI systems are of interest because they yield combinatorial min-max theorems.
Moreover, integrality of the right-hand side of a TDI system implies that the underlying polytope is integer~\cite{EDMONDS1977185}. 


\medskip

Now, we recall the literature on co-2-plex polytopes.
The following trivial inequalities are facet-defining by the fact that each singleton and each pair of vertices is a co-2-plex~\cite{bala}.

\begin{obs}[\cite{bala}]\label{obs:facettetrivialco2plex}
	Given a graph $G=(V,E)$, the following inequalities define facets of $\mathcal{P}(G)$:
	\begin{equation}
		-x_v \le 0 \ \text{ for all }v \in V, \label{iq:trivNatural0}
	\end{equation}
	\begin{equation}
		x_v \le 1 \ \text{ for all }v \in V. \label{iq:trivNatural1}
	\end{equation}
\end{obs}

Balasundaram et al.~\cite{bala} proved that a co-2-plex can intersect at most two times any 2-plex. This remark naturally yields valid inequalities for the co-2-plex polytope, hereafter called \emph{2-plex inequalities}.

\begin{theorem}[\cite{bala}]\label{the:iq2plex}
    Given a graph $G$, the family of 2-plex inequalities:
         \begin{equation}
        x(K) \le 2 \  \text{for all  2-plexes $K$ of $G$} \label{eq:2plex}
        \end{equation}
        are valid for $\mathcal{P}(G)$.
        Moreover, the 2-plex inequality~\eqref{eq:2plex} associated with any co-2-plex $K$ of $G$ is facet-defining for $\mathcal{P}(G)$ if and only if $K$ is a maximal 2-plex. 
\end{theorem}

It turns out that any set of vertices intersecting at most 2 times each 2-plex is a co-2-plex. For this reason, Inequalities~\eqref{eq:2plex} and binary constraints on $x$ are sufficient to characterize co-2-plex incidence vectors of a graph. A natural question is when replacing integrality constraints with $0,1$ bounds on the variables defines an integer polytope, equivalently: for which graphs do Inequalities~\eqref{iq:trivNatural0}--\eqref{eq:2plex} define an integer polytope.
McClosky et al.~\cite{pol} answered this question as stated by the following.

\begin{theorem}[\cite{pol}]\label{the:mclocksky}
    The co-2-plex polytope of a connected graph $G$ is equal to
    \begin{equation}
\{x \in [0,1]^{V} : x \text{ satisfies } \eqref{eq:2plex}  \} \label{pol:mclocksky}
    \end{equation}    
      if and only if $G$ is a 2-plex, a co-2-plex, a path, or a hole whose length is a multiple of 3.
\end{theorem}

When a hole has a length which is not a multiple of 3, a rank facet of the following form appears in its co-2-plex polytope.

\begin{theorem}[\cite{pol}] \label{the:mclockskyhole}
If $G=(V,E)$ is a hole such that $|V| \mod 3 \neq 0$, then the following inequality is facet-defining for $\mathcal{P}(G)$:
\begin{equation}
	x(V) \le \left\lfloor \frac{2|V|}{3} \right\rfloor. \label{iq:hole}
\end{equation}

\end{theorem}

Finally, we recall a description of the orthogonal projection of a polyhedron onto a subspace. 

\begin{theorem}[\cite{cocozaEF}] \label{the:fourier}
Given a  $q\times n$ matrix $A$ and a $q\times m $ matrix $B$, the following holds: 
    $$ \{ x \in \mathds{R}^{n} : \exists y \in \mathds{R}^m ~ Ax + By \le c \} = \{ x \in \mathds{R}^n : u^\top Ax \le u^\top c,\ \forall u \in \mathds{R}^q_+  \text{ satisfying } u^\top B = 0 \}.$$
\end{theorem}

By Theorem~\ref{the:fourier}, and since $\mathcal{C}_B=\{u \in \mathds{R}^q_+: u^\top B = 0\}$ is a cone, finding a set of generators of $\mathcal{C}_B$ yields a finite superset of the facets of the projection. Note that Theorem~\ref{the:fourier} implies that the orthogonal projection of a polyhedron is also a polyhedron.

\subsection{A natural space integer linear  formulation}\label{sec:LitteratureReviewILP}

In general, one does not want to enumerate Inequalities~\eqref{eq:cliqueStableSetPolytope} as they come in an exponential number. Hopefully, to obtain a valid ILP for finding maximum $w$ weighted stable sets, it is sufficient to consider a subset of cliques.

\begin{obs}\label{obs:formStableSet}
	Given a graph $G=(V,E)$ and a set of cliques $\mathcal{K}$, we have $\bigcup_{K \in \mathcal{K}} E(K) = E$ if and only if the incidence vectors of $G$'s stable sets are characterized by the elements of $\{0,1\}^{V}$ satisfying Inequalities~\eqref{eq:cliqueStableSetPolytope} restricted to $\mathcal{K}$.
\end{obs}

In particular, if $\mathcal{K}$ is the set of pairs of adjacent vertices of $G$, Observation~\ref{obs:formStableSet} yields the \emph{edge formulation}, whose linear relaxation has been extensively studied~\cite{BIANCHI202370,Francesco, circuit,sassano}. 

Note that, if $\mathcal{K}$ is the set of maximal cliques of $G$ then, $\bigcup_{K \in \mathcal{K}} E(K) = E$ holds. Moreover, if $K$ is not maximal in Inequality~\eqref{eq:cliqueStableSetPolytope}, then the associated inequality is dominated by a combination of trivial and maximal clique inequalities.

The standard ILP for solving the maximum $w$-weighted co-$k$-plex is as follows (originally, it has been written for finding $k$-plexes, by complementing the graph, one obtains the following).

\begin{theorem}[\cite{bala}]
	The following formulation $\mathcal{N}_k(G)$ is valid for finding a maximum $w$-weighted co-$k$-plex of a graph $G=(V,E)$:

	\begin{equation}
		\max \{ w^\top x : x \in \{0,1 \}^V \text{ s.t. } x(N(u)) + (|N(u)|-k+1) x_u  \le  |N(u)|, \text{ for all $u \in V$}\}.
	\label{form:baseline}	
\end{equation}	
	   
\end{theorem}

The linear inequalities in~\eqref{form:baseline} correspond to the fact that in any co-$k$-plex $S$, if $v$ belongs to $S$, then at most $k-1$ of its neighbors belong to $S$. 

\medskip

For the case $k=2$, Balasundaram et al.~\cite{bala} devise a branch-and-cut algorithm where Formulation $\mathcal{N}_2(G)$ defined in~\eqref{form:baseline} is reinforced with 2-plex Inequalities~\eqref{eq:2plex}. These strengthening inequalities are heuristically separated as follows. Given a current continuous solution \(\bar x \in \mathbb{R}^V\), vertices \(v \in V\) are ordered following the decreasing value of their associated variable \(\bar x_v\). Following this order, a clique $C$ is constructed by iteratively adding each vertex that is adjacent to every vertex of $C$. The vertices of $V \setminus C$ are then ordered following their decreasing degree in $G[V \setminus C]$. Following that order, each vertex is iteratively added to $C$ if its union with $C$ is a 2-plex. The 2-plex inequality~\eqref{eq:2plex} associated with $C$ is added to the current relaxation if it is violated by $\bar x$.

\subsection{Preprocessing for finding maximum cardinality co-2-plexes \label{sec:preprocess}}

Gao et. al in~\cite{doi:10.1142/S0218213019500155} show that given a lower bound $b$ for the size of the largest co-$k$-plex of a graph $G=(V,E)$, if a vertex has a neighborhood of size greater than or equal to $|V|-b+k$, then this vertex belongs to no largest co-$k$-plex of $G$. This permits to safe removal the vertices verifying this property and doing so iteratively.
Note that the diameter of a $k$-plex is bounded by $k$. This remark led Gao et. al~\cite{doi:10.1142/S0218213019500155} to decompose an instance $G=(V,E)$ into $|V|$ smaller instances:
instead of computing the maximal size co-$k$-plex problem on $G$, they compute for each vertex $v \in V$ the subgraph of $G$ induced by the vertices at distance at most $k$ from $v$ in $\overline{G}$. Finally, they solve the maximum co-$k$-plex problem on each of the $|V|$ obtained instances. Combining this preprocessing and decomposition has been experimentally successful in solving the problem on hard instances. 

\section{Polyhedral contributions}\label{sec:polyhedralContributions}

We first give an orthogonal extended formulation for the co-2-plex polytope of a contraction perfect graph $G=(V,E)$ having $|V| + |E|$ variables. More precisely, we give a polyhedral description of the polytope:
\begin{equation*}
\mathcal{P}_{x,y}(G) = \text{conv}\{ (\chi^S,\zeta^{E(S)}) \in \{0,1 \}^{V\times E} \mid S \text{ is a co-2-plex of $G$ }  \},
\end{equation*}
When $G$ is contraction perfect. By definition, $\mathcal{P}(G) = \text{proj}_x(\mathcal{P}_{x,y}(G))$. Then, we show that the extended formulation is compact when the graph is chordal. We finally project the extended formulation into the natural space when the graph is a tree.


\subsection{An extended linear description for contraction perfect graphs}

Given a graph $G=(V,E)$, $W \subseteq V$, and $F\subseteq E$, we say that $[W,F]$ is an \emph{utter clique} of $G$ if $W\cup F $ is a clique of $u(G)$.
Then, $W$ is a clique of $G$.
An utter clique $[W,F]$ is \emph{maximal} if no element can be added to $W$ or $F$ such that it remains an utter clique. 
In this case, we have $E(W)\subseteq F$.

\begin{theorem}\label{the:contractionperfectIntegralPol}
	A graph $G = (V,E)$ is contraction perfect if and only if 
    $\mathcal{P}_{x,y}(G)$ is the set of points $(x,y) \in \mathbb{R}^V \times \mathbb{R}^E$ satisfying:
	\begin{align}
		x(W) + y(F \cap E(V\setminus W))  -y(E(W))  \le  1 && \begin{array}{r}\text{ for all maximal utter cliques }\\  \text{ $[W,F]$ of } G,\end{array}\label{iq:cliquestorng}\\
		y(\delta(v)) \le x_v && \text{ for all }v \in V,\,\,\,\label{iq:uk_delta}\\
		-y_e \le 0 && \forall e \in E.\,\,\, \label{eq:triviale}
	\end{align}
\end{theorem}
\noindent\begin{proof}
	We prove this result by a variable change on a linear system describing the stable set polytope of $u(G)$.
As $u(G)$ has vertex set $V \cup E$, we describe its stable set polytope with variables $z\in \mathds{R}^V$ and $y\in \mathds{R}^E$.

	By Theorem~\ref{the:matching} and Corollary~\ref{cor:perfect}, a graph $G=(V,E)$ is contraction perfect if and only if the stable set polytope of $u(G)$ is described by Inequalities~\eqref{eq:cliqueStableSetPolytope} and~\eqref{eq:trivialPositivity}, that is, by the following system:
	\begin{align}
		z(W) +  y(F)    \le  1 &&\begin{array}{r}\text{ for all maximal utter cliques }\\ \text{ $[W,F]$ of } G,\end{array}  \label{ineq:tdi1}\\
		-z_u  \le  0 && \text{ for all }u \in V,\,\,\, \label{ineq:tdi2}\\
		-y_{e}  \le   0 && \text{ for all }e \in E.\,\,\,\label{ineq:tdi3}
	\end{align}	

	Given a stable set $S\cup M $ of $u(G)$ with $S \subseteq V$ and $ M \subseteq E$, the corresponding co-2-plex of $G$ contains a vertex $u$ either if $u \in S$ or if $uv \in M$.
    Moreover, if $uv\in M$, then $\delta(u)\cap M = \{uv\}$. This remark naturally imposes the following variable change:
	 $$
	\phi = \left\{\begin{array}{rcll}
	(z,y) \in [0,1]^{V\times E} &\rightarrow & (x,y) \in [0,1]^{V\times E} \\
	x_u &=& z_u + y(\delta(u)) & \forall u \in V.	
	\end{array}\right.$$

Applying $\phi$ to system \eqref{ineq:tdi1}--\eqref{ineq:tdi3} yields \eqref{iq:cliquestorng}--\eqref{eq:triviale} by the following:
\begin{itemize}
	\item Inequalities~\eqref{ineq:tdi3} immediatly yield Inequalities~\eqref{eq:triviale}.
	\item Inequalities~\eqref{ineq:tdi2} immediatly yield Inequalities~\eqref{iq:uk_delta}.
	\item Inequalities~\eqref{ineq:tdi1} yield Inequalities~\eqref{iq:cliquestorng} as follows: given a maximal utter clique $[W,F]$, $F$ is partitionned into $F_i, \ i=0,1,2$ where $F_i$ is the subset of $F$ having $i$ endpoints in $W$ that is by maximality~: $F_2 = E(W)$, $F_1 = \delta(W)$ and $F_0 = F\cap E(V\setminus W)$. This yields the following equalities:
	$$z(W) + y(F) = x(W)- y(F_1)-2y(F_2) + y(F_0) + y(F_1) + y(F_2)  = x(W) + y(F \cap E(V\setminus W))  -y(E(W)).$$ 
\end{itemize}

Note that Inequalities~\eqref{iq:cliquestorng}--\eqref{eq:triviale}  are in one-to-one correspondence with Inequalities~\eqref{ineq:tdi1}--\eqref{ineq:tdi3}, hence, $\phi(z,y)$ and $(z,y)$ satisfy the same set of inequalities up to the application of $\phi$. For this reason, $\phi(z,y)$ is an extreme point of $\{(x,y) \text{ satisfying \eqref{iq:cliquestorng}--\eqref{eq:triviale}}\}$ if and only if $(z,y)$ is an extreme point of $\{ (z,y) \text{ satisfying \eqref{ineq:tdi1}--\eqref{ineq:tdi3}}\}$. Moreover, $\phi(z,y)$ is integer if and only if $(z,y)$ is integer. This proves that the integrality of both polytopes coincide, proving the equivalence between contraction perfectness of a graph and integrality of the associated system \eqref{iq:cliquestorng}--\eqref{eq:triviale}.

By construction, if $(x,y)$ satisfying \eqref{iq:cliquestorng}--\eqref{eq:triviale} is integer, then $x$ is the incidence vector of a co-2-plex $S$ of~$G$ and $y$ is the incidence vector of $E(S)$.
\end{proof}


Inequalities~\eqref{iq:cliquestorng} stem from clique Inequalities~\eqref{eq:cliqueStableSetPolytope} for the stable set polytope. Similarly to clique inequalities associated with non-maximal cliques that are redundant, we have the following.
\begin{obs}\label{obs:dominatedUtterCliques}
	Given a non maximal utter clique $[W,F]$, the inequality $x(W) + y(F\cap E(V\setminus W)) -y(E(W)) \le 1$ is dominated with respect to Inequalities~\eqref{iq:cliquestorng}--\eqref{eq:triviale}.
\end{obs}
\noindent\begin{proof}
	Let $[W',F']$ be a maximal utter clique such that $W \subseteq W'$ and $F \subseteq F'$. Summing up Inequality~\eqref{iq:cliquestorng} associated with $[W',F']$ with Inequalities~\eqref{iq:uk_delta} associated with $W' \setminus W$ and Inequalities~\eqref{eq:triviale} associated with $E(W' \setminus W) \setminus F$ and $E(V\setminus W')\cap (F' \setminus F)$ yields inequality~\eqref{iq:cliquestorng} associated with $[W,F]$.
\end{proof}

Corollary~\ref{cor:perfect} also yields a characterization of contraction perfectness in terms of TDI-ness of a linear system.

\begin{corollary}\label{cor:tdi}
	A graph $G$ is contraction perfect if and only if \eqref{iq:cliquestorng}--\eqref{eq:triviale} form a TDI system.
\end{corollary}
\noindent\begin{proof}
	By Theorems~\ref{the:matching} and~\ref{the:chva}, a graph $G$ is contraction perfect if and only if \eqref{ineq:tdi1}--\eqref{ineq:tdi3} form a TDI system.
	Suppose this system to be TDI.
	Then, adding the redundant constraints $z_u+y(\delta(u))\geq 0$, for all $u\in V$ yields a TDI system again.
	Moreover, adding and removing slack variables also preserve TDIness~\cite{COOK198331}.
	Introducing a slack variable $-x_u \leq 0$ for each of these new constraints yields the TDI system \eqref{ineq:tdi1}--\eqref{ineq:tdi3}, $z_u+y(\delta(u))= x_u$, $-x_u\leq 0$, for all $u\in V$.
	
	Replacing each $z_u$ in \eqref{ineq:tdi1} by $z_u = x_u - y(\delta(u))$ yields \eqref{iq:cliquestorng}.
	Afterwards, observe that $z_u \geq 0$ acts as a slack variable with respect to~\eqref{iq:uk_delta}.
	In particular, removing $z_u$ preserves TDIness, and hence $y\ge \mathbb{0},$ \eqref{iq:cliquestorng}, and~\eqref{iq:uk_delta} is a TDI system.
	
	The observation that $z_u$ acts as a slack variable makes it possible to backtrack the proof of the two previous paragraphs, up to the addition of $z_u+y(\delta(u))\geq 0$.
	Since these are non-negative integer combinations of~\eqref{ineq:tdi2} and~\eqref{ineq:tdi3}, their deletion preserves TDIness, yielding the desired equivalence.
\end{proof}

\subsection{A compact extended formulation for chordal graphs}

For contraction perfect graphs, the extended formulation given by \eqref{iq:cliquestorng}-\eqref{eq:triviale} is not compact since the number of maximal utter cliques of a contraction perfect graph may be exponential.
We show in this section that the formulation becomes compact when considering chordal graphs. For this, we show that Inequalities~\eqref{iq:cliquestorng} correspond in chordal graphs to the following set of inequalities:



\begin{equation}
	x(K) -y(E(K))  \le  1   \text{ for all maximal cliques } K \text{ of } G.\label{eq:cliquechordal}\\
\end{equation}

The proof of Theorem~\ref{the:charachordal} uses the following characterization of utter graphs for chordal graphs.
\begin{theorem}[\cite{dupontbouillard2024contractions}]\label{thm:GchordalIffuG}
    A graph is chordal if and only if its utter graph is.
\end{theorem}

\begin{theorem}\label{the:charachordal}
	A graph $G$ is chordal if and only if $\mathcal{P}_{x,y}(G)=\{(x,y): \eqref{iq:uk_delta}, \eqref{eq:triviale}, \text{ and}~\eqref{eq:cliquechordal}\}$.
\end{theorem}
\noindent\begin{proof}
Recall that if $[W,F]$ is a maximal utter clique, then $W$ is a clique and $E(W)\subseteq F$.

	$(\Rightarrow)$ Suppose that $G=(V,E)$ is chordal. Chordal graphs are contraction perfect~\cite{dupontbouillard2024contractions}.
    Thus, by Theorem~\ref{the:contractionperfectIntegralPol}, it is sufficient to show that every maximal utter clique $[W,F]$ of $G$ satisfies $F' = F\cap E(V\setminus W) = \emptyset$.
	By contradiction, suppose that there exists an edge $uv$ in $F'$. 
    In particular, neither $u$ nor $v$  belongs to $W$. 
    By maximality, there exists at least one vertex $w_u \in W\cup F$ (resp. $w_v$) that is non-adjacent to $u$ (resp. $v$) in $u(G)$.
    Let $x_{uv}$ be the vertex of $u(G)$ corresponding to $uv$.
	We have $w_u\neq w_v$, as otherwise $x_{uv}$ would be non-adjacent to $w_u$ in $u(G)$, contradicting the fact that $W\cup F$ is a clique of $u(G)$. Similarly, $w_uv$ and $w_vu$ are edges of $u(G)$. But then, $w_v, u, v, w_u$ is a hole of $u(G)$, implying that $G$ is not chordal by Theorem~\ref{thm:GchordalIffuG}, a contradiction.
\medskip
    
	($\Leftarrow$) 
    If $G$ is not contraction perfect,  we have:  
    $$\mathcal{P}_{x,y}(G) \subsetneq \{(x,y): \eqref{iq:cliquestorng}\text{--}\eqref{eq:triviale}\}\subseteq\{(x,y): \eqref{iq:uk_delta}, \eqref{eq:triviale}, \text{ and}~\eqref{eq:cliquechordal}\}$$
    where the strict inclusion comes from Theorem~\ref{the:contractionperfectIntegralPol} and the other inclusion from Observation~\ref{obs:dominatedUtterCliques} and the definition of~\eqref{eq:cliquechordal}.

If $G$ is contraction perfect but not chordal, it contains a hole of size 4, say $(u,v,w,z)$. Let $K_1$, $K_2$ and $K_3$ be three maximal cliques of $G$ satisfying $K_1 \cap \{u,v,w,z\} = \{u,z\}$, $K_2 \cap \{u,v,w,z\} = \{u,v\}$ and $K_3 \cap \{u,v,w,z\} = \{v,w\}$. 
	
	Let $(x^*,y^*)$ be a solution of \eqref{iq:uk_delta}, \eqref{eq:triviale}, \eqref{eq:cliquechordal} such that $x_u^* = x_v^* = x_z^* = x_w^* = y_{zw}^*=\frac{1}{2}$ and all other components are 0. $(x^*,y^*)$ satisfies with equality the clique inequalities associated with $K_1$, $K_2$ and $K_3$, Inequalities~\eqref{iq:uk_delta} associated with each vertex but $u$ and $v$, and Inequalities~\eqref{eq:triviale} associated with every edge except $vw$. These $3 + (|V|-2 ) + (|E|-1) = |V|+ |E|$ inequalities are linearly independent so $(x^*,y^*)$ is a fractional extreme point of $\{ (x,y) :\eqref{iq:uk_delta},~\eqref{eq:triviale}, and~\eqref{eq:cliquechordal} \}$, which ends our proof.
\end{proof}

The proof of Theorem~\ref{the:charachordal} implies that the maximal utter cliques of a chordal graph are in bijection with its maximal cliques. Moreover, as the number of cliques of a chordal graph is linear~\cite{chordal}, the number of Inequalities~\eqref{eq:cliquechordal} is polynomial, which yields the following. 

\begin{corollary}
 There is a compact description of $\mathcal{P}_{x,y}(G)$ in chordal graphs.
\end{corollary}

\subsection{Co-2-plex polytopes in the natural variable space}

In this section, we start with elementary results on $\mathcal{P}(G)$.
Then, we use Theorem~\ref{the:fourier} to get rid of $y$ variables in the linear system \eqref{iq:uk_delta}, \eqref{eq:triviale}, and~\eqref{eq:cliquechordal}.
This yields a linear system describing the co-2-plex polytopes of trees in the natural variable space. It turns out that the latter system is not only integer for trees but also for holes of length multiple of 3. We show that these graphs characterize the integrality of this system among the graphs containing no true twins. 
Having no true twins is a reasonable hypothesis, as removing an edge between true twins does not change the set of co-2-plexes.

\medskip

The \emph{support} of an inequality $ax\le \alpha$ is the graph induced by the set of vertices $u$ satisfying $a_u \neq 0.$ 

\begin{theorem}\label{the:lifting}
	Given a graph $G=(V,E)$ and a facet-defining inequality $ax\le \alpha$ of $\mathcal{P}(G)$, the following inequality defines a facet of $\mathcal{P}(G')$, where $G'$ is obtained from $G$ by adding a vertex $w$ complete to the vertices of the support $H$ of $ax\le \alpha $:
	\begin{equation}
		ax + (\alpha - \max_{v\in V(H)} a_v)x_w \le \alpha. \label{iq:lift}\end{equation}
\end{theorem}
\noindent\begin{proof}
	The validity of Inequality~\eqref{iq:lift} for $\mathcal{P}(G')$ comes from the fact that any co-2-plex containing $w$ contains at most one vertex of $H$, and from the validity of $ax\le \alpha$ for $\mathcal{P}(G)$.
	As $ax\le \alpha$ is facet defining for $\mathcal{P}(G)$,
	it is satisfied with equality for some affinely independent points $\chi^{S_i}, \ i\in \{1,\dots,|V| \}$ such that each $S_i$ is a co-2-plex of $G$.
	Each $S_i$ is also a co-2-plex of $G'$, and its incidence vector satisfies Inequality~\eqref{iq:lift} with equality. Finally, given $u = \textrm{argmax}_{v\in V(H)}(a_v)$, we have that $\chi^{\{ u,w \}}$, $\chi^{S_1}$, $\dots$ $\chi^{S_{|V|}}$ are $|V|+1$ affinely independent vectors satisfying Inequality~\eqref{iq:lift} with equality.
\end{proof}

The inequalities in~\eqref{form:baseline} can be generalized as follows.

\begin{theorem}\label{the:facettes}
	Given a graph $G=(V,E)$, a vertex $w \in V$ and a subset of its neighborhood $W$, 
	\begin{equation}
	x(W) + (\alpha_2(G[W])-1) x_w  \le \alpha_2(G[W]) \label{iq:lifting}
	\end{equation}
	defines a facet of $\mathcal{P}(G)$ if and only if the following hold: 
	\begin{enumerate}[i)]
		\item no vertex $u\in V\setminus (W \cup \{w\})$ is complete to $W$,
		\item $\forall v \in N(w) \setminus W$, $\alpha_2 (G[W \cup v] ) = \alpha_2 (G[W]) + 1$.
	\end{enumerate}
\end{theorem}
\noindent\begin{proof}
	$(\Leftarrow)$
	The validity of Inequalities~\eqref{iq:lifting} is straightforward.
	Note that if $W = \{u \}$, the obtained inequality is $x_u \le 1$, which defines a facet by Observation~\ref{obs:facettetrivialco2plex}. 
	We prove the remaining cases by maximality. Take $w \in V$, $W \subseteq N(w)$ of size at least 2 and consider the inequality $x(W) + (\alpha_2(W)-1)x_w   \le  \alpha_2(W)$ and the face $F$ of $\mathcal{P}(G)$ that it defines. Suppose that there exists a face $F'$ of $\mathcal{P}(G)$ containing $F$ and described by $ax \le \alpha$.

    \begin{itemize}
        \item 
	\textbf{Claim 1.} $a_u = 0$, for all $u \in V{\setminus} N[w]$.
    
    By hypothesis \emph{i)}, $\forall u \in V\setminus N[w]$ there exists a vertex $v$ of $W$ non adjacent to $u$, in this case $\chi^{v,w,u}$ and $\chi^{w,v}$ both belong to $F$, and hence belong to $F'$ implying that $a_v + a_w + a_u = a_v + a_w$ and hence $a_u =  0$. 

    \item
	\textbf{Claim 2} $a_u = 0$, for all $u \in N(w) \setminus W$.
    
    By hypothesis $ii)$, there exists $W' \subseteq W$ such that $W' \cup u$ is a co-2-plex of size $\alpha_2(G[W]) +1$. Note that, $\chi^{W'\cup u}$ and $\chi^{W'}$ both belong to~$F$ which implies that $a_u=0$.
	
\item 	\textbf{Claim 3.} $a_u = a_v$, for all $u,v \in W $.

It is obtained by the fact that $\chi^{v,w}$ and $\chi^{u,w}$ both belong to $F$.

    \item
	\textbf{Claim 4} $\alpha = a_u \alpha_2(G[W])$, for all $u \in W$.
    
    Let $W' \subseteq W$ be a co-2-plex of size $\alpha_2(G[W])$, $\chi^W$ belongs to $F$, by Claim 3, we obtain that $\alpha = a_u \alpha_2(G[W])$.
    
	Finally, as $\chi^{w,u} \in F$ , we have $a_w = \alpha - a_u$ for each $u \in W$, we obtain that $ax\le \alpha $ is a multiple of Inequality~\eqref{iq:lifting} which implies that $F = F'$.
    \end{itemize}
	
	$(\Rightarrow)$
	By contradiction, suppose that there exists a vertex $u \in V \setminus (W \cup \{w\})$ complete to $W$, or that there exists $u \in N(w) \setminus W$ such that $\alpha_2 (W \cup u ) = \alpha_2(W)$,  then $x(W) + (\alpha_2(W) -1) x_w + x_u \le \alpha_2(W)$ is valid for $\mathcal{P}(G)$ and dominates Inequality~\eqref{iq:lifting} with respect to nonnegativity constraints.    
\end{proof}


Theorem~\ref{the:facettes} generalizes Corollary~3.7. and Theorem~3.11 of~\cite{pol}. 
In a tree, the neighborhood of each vertex forms a co-2-plex; hence, we obtain the following by Theorem~\ref{the:facettes}.

\begin{corollary}\label{the:facetrees}
	Given a graph $G=(V,E)$, the star inequalities:
	\begin{equation}
		x(W) + (|W|-1)x_w   \le  |W|, \text{ for all $w \in V$ and $W \subseteq N(w)$} \label{iq:tree}
	\end{equation}
are valid for $\mathcal{P}(G)$. Moreover, if $G$ is a tree, these inequalities define facets of $\mathcal{P}(G)$.
\end{corollary}

Let us denote by $\mathcal{T}(G)$ the polytope defined by~\eqref{iq:tree} and $0\le x \le 1$.
We give structural results towards a characterization of the integrality of $\mathcal{T}(G)$. For this, we consider the following set of inequalities.

\begin{obs}\label{obs:iqedge}
	The following set of inequalities is valid for $\mathcal{P}_{x,y}(G)$:
	\begin{equation}
		x_u + x_v - y_{uv} \le 1 \text{ for all $uv \in E$}.  \label{iq:edgesExtended}
	\end{equation}
\end{obs}
\noindent\begin{proof}
These inequalities are obtained by Inequalities~\eqref{iq:cliquestorng} associated with the following utter cliques $\{ [\{u,v\}, \delta(u) \cup \delta(v)] : \forall uv \in  E \}$. If the latter is not maximal, Observation~\ref{obs:dominatedUtterCliques} yields the validity.
\end{proof}

\begin{lemma}\label{the: firstTreecharac}
  Given a graph $G=(V,E)$, the following holds:
 $$  \mathcal{T}(G)= \{x : \exists y \ \text{s.t. }(x,y) \text{ satisfies  \eqref{iq:uk_delta}, \eqref{eq:triviale}, \text{and}~\eqref{iq:edgesExtended}} \}.$$
\end{lemma}
\noindent\begin{proof}
	We will apply Theorem~\ref{the:fourier} to compute $\proj_x\{(x,y):\text{ $(x,y)$ satisfies  \eqref{iq:uk_delta},~\eqref{eq:triviale},~\eqref{iq:edgesExtended}} \}$. We first provide a matricial representation of  the linear system \eqref{iq:uk_delta}, \eqref{eq:triviale}, \eqref{iq:edgesExtended} of the form $\{(x,y) \in \mathds{R}^{V} \times \mathds{R}^E: Ax + By \le c \}$ where: 
$$A= \begin{pmatrix} M \\
	-\mathds{I}_{|V|} \\
	\mathbb{0}_{|V|\times |E|} \end{pmatrix}\!\!,
B = \begin{pmatrix} -\mathds{I}_{|E|} \\ M^\top \\ -\mathds{I}_{|E|} \end{pmatrix}\!\!,
c = \begin{pmatrix}
	\mathbb{1}_{|E|}\\
	\mathbb{0}_{|V|} \\
	\mathbb{0}_{|E|}
\end{pmatrix}\!\!,$$

where $M$ is the $E\times V$ edge-vertex incidence matrix of $G=(V,E)$, namely $M_e^v = 1$ if $e \in \delta(v)$ and $0$ otherwise.
For every $v\in V$ and $F\subseteq \delta(v)$, we define:
$$
g_{v,F} = \begin{pmatrix}
	\zeta^F \\
	\chi^v \\
	\zeta^{\delta(v)\setminus F} 
\end{pmatrix}\!\!,
$$
and let $\cal{G}$ be the set of all $g_{v,F}$'s.

Let us prove that the cone $C=\{u\in \mathds{R}^{|E|\times |V| \times |E|}_+: u^\top B = 0\}$ is generated by $\cal G$.
First, note that $\mathcal{G} \subseteq C$.
Now, let $u$ be a nonzero vector in $C$.
Denoting $u^\top =(p^\top,q^\top,r^\top)$ with $(p,q,r)\in \mathds{R}^{|E|\times |V| \times |E|}_+$, the equalities $u^\top B = 0$ are $q_v + q_w = p_{vw}+r_{vw}$ for all $vw \in E$.
Since $u$ is nonzero and nonnegative, there exists $v\in V$ such that $q_v> 0$.
Let $F=\{e\in\delta(v): p_e >0\}$ and $\rho = \min\{q_v, p_e:e \in F\}$, and note that $r_e \geq \rho$ for all $e\in \delta(v)\setminus F$.
Then, $u' = u - \rho g_{v,F}$ belongs to $C$, satisfies $u'\leq u$, and has at least one more zero component than $u$.
Repeating this decomposes $u$ into a nonnegative combination of vectors in $\cal G$, hence $C = \cone (\cal G)$.

The latter and Theorem~\ref{the:fourier} imply that $\proj_x\{ (x,y): \text{$(x,y)$ satisfy \eqref{iq:uk_delta}, \eqref{eq:triviale},~\eqref{iq:edgesExtended}} \}$  is obtained by computing $g^\top A x\leq g^\top c$ for every $g\in \cal G$, that is, $g_{v,F}^\top A x\leq g_{v,F}^\top c$ for every $v\in V$ and $F\subseteq \delta(v)$.
Note that the sets of edges $F\subseteq \delta(v)$ are in one-to-one correspondence with the subsets of vertices $W\subseteq N(w)$.
For $F\subseteq \delta(v)$, denoting by $W$ the subset of $N(v)$ such that $F=\{vw: w\in W\}$, the inequality $g_{v,F}^\top A x\leq g_{v,F}^\top c$ becomes: 
$x(W) + (|W|-1)x_v \le  |W|,$
which ends the proof.
\end{proof}

  As trees are chordal and their maximal cliques are edges, we obtain the following as a corollary of Theorem~\ref{the:charachordal} and Lemma~\ref{the: firstTreecharac}.

\begin{corollary}\label{the: treecharac}
The co-2-plex polytope of a tree $G$ is equal to $\mathcal{T}(G)$.	
\end{corollary}

It turns out that $\mathcal{T}(G)$ coincides with~\eqref{pol:mclocksky} when $G$ is a hole of length multiple of 3 or a triangle. This shows that Corollary~\ref{the: treecharac} is not a characterization. Towards such a result, we first need the following three technical results.

\begin{proposition}\label{prop:twins}
	Let $G^f$ be a graph having two false twins $u$ and $v$, and let $G^t$ be obtained by adding an edge $uv$ to $G^f$. The set of co-2-plexes of $G^t$ is equal to the set of co-2-plexes of $G^f$.  
\end{proposition}
\noindent\begin{proof}
	Since $E(G^t)= E(G^f)\cup \{uv\}$, the co-2-plexes of $G^t$ are co-2-plexes of $G^f$.
    Conversely, let $S$ be a co-2-plex of $G^f$ containing $u$ and $v$. As $u$ and  $v$ are false twins, every vertex in $S \cap N(u)$ is also in $S \cap N(v)$. By definition of co-2-plexes, $S \cap N(u) = S \cap N(v) = \emptyset$, which implies that $S$ is also a co-2-plex of $G^t$.
\end{proof}

Proposition~\ref{prop:twins} implies that adding an edge between two false twins $u$ and $v$ of $G$ yields a graph $G'$ whose co-2-plex polytope is equal to the one of $G$, this remark permits us to focus on graphs with no true twins. 

\medskip

Given a matrix $A$ and a set of columns $J$, we denote by $A^J$ the submatrix of $A$ induced by the columns of $J$.

\begin{obs}\label{obs:subgraphs}
	If $\mathcal{P}(G) = \{x \in \mathds{R}^V : Ax\le b \}$, then for every induced subgraph $G'$ of $G$, the following holds: \begin{equation}
		\mathcal{P}(G') = \{x \in \mathds{R}^{V(G')} : A^{V(G')} x \le b\}. \label{eq:proj}
	\end{equation}
\end{obs}
\noindent\begin{proof}
	The co-2-plexes of $G'$ are exactly the co-2-plexes of $G$ containing no element of $V(G) \setminus V(G')$. For this reason, $\mathcal{P}(G') = \textrm{conv}\{ x' \in \{ 0,1\}^{V(G')}:   (x',\mathbb{0}) \in \mathcal{P}(G)    \}$. As the equalities $x_v =0$ for all $ v \in V$ define faces of $\mathcal{P}(G)$ which is itself integer, we have that $\{x \in \mathcal{P}(G) : x_v = 0, \text{ for all } v \in V(G) \setminus V(G') \} $ is integer. Therefore, Equality~\eqref{eq:proj} holds. 
\end{proof}

Note that Observation~\ref{obs:subgraphs} also holds for the co-$k$-plex polytope with $k\neq 2$.
Observation~\ref{obs:subgraphs} is used to focus on graphs not containing some particular induced subgraphs.

\medskip

Finally, we obtain the characterization of when $\mathcal{T}(G)$ is the co-2-plex polytope of $G$.

\begin{theorem}\label{the:characpolytope}
	Let $G$ be a graph with no true twins. 
    Then, $\mathcal{P}(G) = \mathcal{T}(G)$ if and only if $G$ is a tree or a hole of length multiple of 3.
\end{theorem}
\noindent\begin{proof}
	$(\Leftarrow)$ If $G$ is a tree, then $\mathcal{P}(G) = \mathcal{T}(G)$ by Corollary~\ref{the: treecharac}.
    If $G$ is a hole of length multiple of 3, then the maximal $2$-plexes of $G$ are the sets of three consecutive vertices, hence \eqref{eq:2plex} and \eqref{iq:tree} coincide, and the result follows from Theorem~\ref{the:mclocksky}.

    \medskip
    
	$(\Rightarrow)$ We first prove the following.    
    
\begin{claim}\label{lem:forbidden}
	If $\mathcal{P}(G)= \mathcal{T}(G)$, then $G$ contains neither 2-plexes of size 4, nor holes whose lengths are not multiples of 3.
\end{claim}
\noindent\begin{proof}
	Using Observation~\ref{obs:subgraphs}, it is sufficient to exhibit a facet defining inequality for $\mathcal{P}(G[H])$ that is not obtained by combining a facet defining inequality of $\mathcal{T}(G)$ with $x_v = 0$ for all $v \in V \setminus H$, and do so for the cases where 	$H$ induces a 2-plex of size 4 or a hole whose length is not multiple of 3.
	
	When $H$ induces a 2-plex of size 4, then $H$ is a maximal $2$-plex of $H$, hence by Theorem~\ref{the:iq2plex}, the corresponding facet-defining inequality for $\mathcal{P}(G[H])$ is $x(H) \le 2$. No Inequality~\eqref{iq:tree} has a support of at least 4 vertices and a right-hand side equal to 2. 
	When $H$ induces a hole whose length is not a multiple of 3, the corresponding facet-defining inequality for $\mathcal{P}(G[H])$ is $x(H) \le \left\lfloor \frac{2|H|}{3} \right\rfloor$ by~\ref{the:mclockskyhole}. 
	\end{proof}

By contradiction, suppose that $G=(V,E)$ is not a tree and not a cycle of length multiple of 3. 
	As $G$ is not a tree, it contains a smallest hole induced by $H$ or a triangle. Let us consider the following three cases.
    
\begin{enumerate}    
    \item\label{case1} $G$ contains a triangle $T = \{t_1,t_2,t_3 \} \subseteq V$. $G$ is not a triangle as it contains no true twins hence, there exists at least one vertex $u \in V\setminus T$ adjacent to $T$. If $u$ is adjacent to two vertices, say $t_1, \ t_2$, then by Claim~\ref{lem:forbidden}, $\mathcal{T}(G)$ is not integer as $G$ contains the 2-plex induced by $\{t_1,t_2,t_3,u\}$.
	If $u$ is adjacent to exactly one vertex, say w.l.o.g $t_1$, then as $t_2$ and $t_3$ are not true twins w.l.o.g. there exists a vertex $v\neq u$ adjacent to $t_2$ and not $t_1$ nor $t_3$  In this case, $x_{u}+ x_{v} + x_{t_1} + x_{t_2} + x_{t_3}\le 3$ is facet defining for $\mathcal{P}(G[\{u,v,t_1,t_2,t_3\}])$ independently of the adjacency of $u$ and $v$ and is not obtained by restricting a facet defining inequality for $\mathcal{T}(G)$ to elements of $\{u,v,t_1,t_2,t_3 \}$. We obtained this facet using JuliaPolyhedra~\cite{legat2023polyhedral}. 
	
	\item\label{case2} $H$'s size is not a multiple of 3. Claim~\ref{lem:forbidden} concludes.
	
	\item  $H$ induces a hole whose size is a multiple of $3$. 
    Let $(v_0, \dots, v_{p-1} )$ be an associated sequence of vertices.
    As $G$ is not a hole, there exists a vertex $u$ of $V\setminus H$ adjacent to some vertices of $H$. 
	
	

    Suppose that $u$ is adjacent to at least two vertices of $H$.
	If two of these vertices are adjacent, then together with $u$ they form a triangle of $G$, a contradiction as in case~\ref{case1}.
	Otherwise, $H\cup\{u\}$ contains a hole whose size is not a multiple of $3$, a contradiction as in case~\ref{case2}.
	
	Thus, $u$ is adjacent to a single vertex of $H$, say $v_1$.
    We exhibit a non-integer extreme point $x^*$ of $\mathcal{T}(G)$ as follows:
	
	$$x^* = \frac{1}{2}\left(\chi^u+ \sum\limits_{\displaystyle \begin{array}{c}
			\scriptstyle i = 1 \\
			\scriptstyle i\mod 3 \neq 1
	\end{array} }^{p-1} \chi^{v_i} \right) + \sum\limits_{\displaystyle \begin{array}{c}
		\scriptstyle i = 1 \\
		 \scriptstyle  i\mod 3 =1
		\end{array} }^{p-1} \chi^{v_i}.$$
	
	First, we show that $x^*$ is inside $\mathcal{T}(G)$; if an inequality~\eqref{iq:tree} associated with $w\in V$ and $W \subseteq N(w)$ is violated by $x^*$, then it can be chosen such that $w \in H\cup\{u\}$ and $W \subseteq N(w) \cap (H \cup \{u\})$, in particular $|W| \le 3$. 
	If $W$ has size two, the corresponding inequality is the sum of three vertices' components, one being adjacent to the two others, which never exceeds 2 over $x^*$. If $W$ has size 3, we immediatly have $w = v_0$ and $W = \{u,v_1,v_p \}$ which yields an inequality satisfied by $x^*$. 
 
  Now, we show that it is an extreme point. The first matrix of Figure~\ref{fig:proofOfcharac} say $M$, yields a set of valid inequalities satisfied with equality by $x^*$. They correspond to Inequalities~\eqref{iq:tree} associated with $v_i$ and $W = \{v_{i-1},v_{i+1} \}$ for $i = 0,\dots,p-2$ considering $v_{-1} = v_{p-1}$, \eqref{iq:tree} associated with $v_0$ and $W = \{u,v_1\}$, and \eqref{eq:triviale} associated with $v_1$. The second matrix say $M'$ is obtained as follows. We have $M_i' = M_i$ for all $i=1,\dots,3$, then, $M_4' = M_4 - M_2$ and finally $M_i' = M_i - M_{i-1}' - M_{i-2} '$ for $i = 5,\dots, p+1$. $M'$ is an upper triangulated matrix with non-zero values on the diagonal, so the $|H|+1$ constraints are linearly independent. Since $x^*$ also satisfies $x^*_v = 0 \ \forall v \in V \setminus (H \cup \{u \})$, it satisfies $|V|$ linearly independent constraints so $x^*$ is a fractionnal extreme point of $\mathcal{T}(G)$. 
\end{enumerate}	
\end{proof}

\begin{figure}
	\centering
	\resizebox{10cm}{!}{
	\begin{tikzpicture}

		\node at(-3.65,2.75) {$u$};
		\node at(-2.82,2.75) {$v_0$};
		\node at(-2,2.75) {$v_1$};
		\node at(-1.2,2.75) {$v_2$};
		\node at(-0.4,2.75) {$v_3$};
		\node at(0.4,2.75) {$v_4$};
		\node at(1.2,2.75) {$v_5$};
		\node at(2,2.75) {$v_6$};
		\node at(3.75,2.75) {$v_{p-1}$};
		
		\node at(5.75,2.1) {$x_u + x_{v_0} + x_{v_1} = 2$};
		\node at(6,1.6) {$ x_{v_0} + x_{v_1} + x_{v_{p-1}} = 2$};
		\node at(4.85,1.15) {$x_{v_1} = 1$};
		\node at(5.75,0.7) {$x_{v_0} + x_{v_1} + x_{v_2} = 2$};
		\node at(5.75,0.3) {$ x_{v_1}+ x_{v_2} + x_{v_3}  = 2$};

		\node at(5.75,-0.1) {$x_{v_3} + x_{v_4} + x_{v_5} = 2$};
		\node at(5.75,-0.5) {$x_{v_4} + x_{v_5} + x_{v_6} = 2$};
		\node at(6.5,-1.4) {$x_{v_{p-4}} + x_{v_{p-3}} + x_{v_{p-2}} = 2$};
		\node at(6.5,-1.8) {$x_{v_{p-3}} + x_{v_{p-2}} + x_{v_{p-1}} = 2$};
		
		\node at (0,0) {$\begin{pmatrix}
				1&1&1& & & & & &\dots& \\
				 &1&1& & & & & &\dots&1\\
				 & &1& & & & & &\dots& \\
				 &1&1&1& & & & &\dots& \\
				 & &1&1&1& & & &\dots& \\
				 & & &1&1&1& & &\dots& \\

				 & & & &1&1&1& &\dots& \\
				\dots&\dots&\dots&\dots&\dots&\dots&\dots&\dots&\dots&\dots\\
				& & & & &\dots&1&1&1& \\
				& & & & &\dots& &1&1&1\\

			\end{pmatrix}$};    
		
		\node at (0,-5)  {$\begin{pmatrix}
				
				1&1&1& & & & & &\dots& \\
				 &1&1& & & & & &\dots&1\\
				 & &1& & & & & &\dots& \\
				 & & &1&& & & &\dots&-1\\
				 & & & &1& & & &\dots&1 \\
				 & & & & &1& & &\dots& \\
				 & & & & & &1& &\dots&-1\\
				 & & & & & & &1&\dots& \\
				\dots&\dots&\dots&\dots&\dots&\dots&\dots&\dots&\dots&\dots\\
				& & & & &\dots& & &1& \\
				& & & & &\dots& & & &-1\\

			\end{pmatrix}$};    
	\end{tikzpicture}}
	\caption{Matrix of constraints satisfied with equality and a scaled version\\ for the proof of Theorem~\ref{the:characpolytope}.}
	\label{fig:proofOfcharac}
\end{figure}

\section{Extended space integer linear formulations}\label{sec:LPformulations}

In this section, we provide different ILPs for the maximum weighted co-2-plex problem. These formulations stem from the bijection between co-2-plexes in a graph and stable sets in its utter graph.

\subsection{Tighter formulations}

A result similar to Observation~\ref{obs:formStableSet} can be deduced using Inequalities~\eqref{iq:cliquestorng} restricted to nonmaximal cliques.
The {\em extended incidence vector} of a co-2-plex $S$ is $(\chi^S,\zeta^{E(S)})$.

\begin{obs}\label{obs:validco2plex}
	Given a set of utter cliques $\mathcal{U}$ of a graph $G$, if $\bigcup_{[W,F] \in \mathcal{U}} E(W)= E$ then $G$'s co-2-plex extended incidence vectors are the elements of $\{0,1 \}^{V}\times\{0,1 \}^{E}$ satisfying Inequalities~\eqref{iq:cliquestorng} associated with~$\mathcal{U}$  and Inequalities~\eqref{iq:uk_delta}.
\end{obs}
\noindent\begin{proof}
	Let $\mathcal{U}$ satisfy the above condition.
	
	As variables associated with $\delta(W)$ vanish when applying $\phi$ in Inequalities~\eqref{ineq:tdi1}, the result is proven by showing that the cliques $\mathcal{K} = \{W\cup F \cup \delta( W) : (W,F) \in \mathcal{U} \}$ of $u(G)$  satisfy the conditions of Observation~\ref{obs:formStableSet}. 
    More precisely, we consider pairs of adjacent vertices $\{u,v\}$ of $u(G)$ and show that they both belong to an element of $\mathcal{K}$. 
    There are four cases.
    \begin{enumerate}
        \item $u,v\in V$.
        Then, by hypothesis there exists an utter clique $[W,F] \in \mathcal{U}$ such that $\{u,v \} \subseteq W$ and hence an element of $\mathcal{K}$ contains both $u$ and $v$.
        
	\item $u \in V$ and $v \in E$ (without loss of generality). Then, $v$ is an edge $wt$ of $G$. By construction, at least one of $w$ or $t$ is adjacent to $u$, say $w$. By hypothesis, there exists a utter clique $[W,F] \in \mathcal{U}$ such that $W$ contains both $u$ and $w$.
    Moreover, $wt$ belong to $E(W)$ or $\delta (W)$ hence, there exists an element of $\mathcal{K}$ containing both $u$ and $wt = v$.

     \item $u,v\in E$ are adjacent in $G$.
     Then, let $w$ be their common vertex.
     Given any utter clique $[W,F] \in \mathcal{U}$ with $w\in W$, we have that $\{u,v\} \subseteq E(W) \cup \delta (W) $, thus an element of $\mathcal{K}$ contains both $u$ and $v$.

     \item  $u=w_1z_1, v= w_2z_2 \in E$ are adjacent by contraction.
     Then, without loss of generality, $w_1w_2$ belongs to $E$. There exists $[W,F]\in\mathcal{U}$ such that $W$ contains both $w_1$ and $w_2$, and then $w_1z_1$ and $w_2z_2$ both belong to $E(W) \cup \delta (W)$.  
    \end{enumerate}
\end{proof}

 Using Observation~\ref{obs:validco2plex}, we obtain the following. 

\begin{obs}
	The ILP $\mathcal{E}(G) =  \max \{w^\top x :  (x,y) \in \{0,1 \}^{V\times E} \text{ satisfies }\eqref{iq:uk_delta}, \eqref{eq:triviale},\eqref{iq:edgesExtended} \}$ is valid for finding maximum $w$-weighted co-2-plexes.
\end{obs}

Again, using Observation~\ref{obs:dominatedUtterCliques}, adding binary constraints and a linear objective function on systems \eqref{iq:cliquestorng}--\eqref{eq:triviale} and~\eqref{iq:uk_delta}, \eqref{eq:triviale},~\eqref{eq:cliquechordal}
also yields two valid ILPs for the maximum co-2-plex problem.

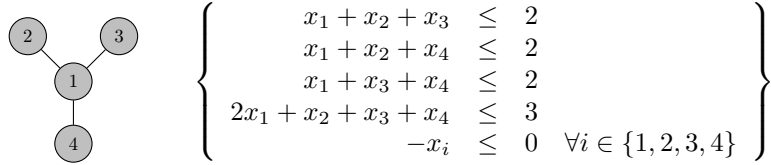
\begin{figure}[h]
	\centering
	\begin{tikzpicture}[state/.style={circle, draw, minimum size=0.7cm},scale = 0.30]
		\node[state,draw,circle,fill=gray!50,scale = 0.7] (1)at(-2,2) {1};
		\node[state,draw,circle,fill=gray!50,scale = 0.7] (2)at(-4,4) {2};
		\node[state,draw,circle,fill=gray!50,scale = 0.7] (3)at(0,4) {3};
		\node[state,draw,circle,fill=gray!50,scale = 0.7] (4)at(-2,-.75) {4};
		\node[opacity=0, text opacity  = 1] (5)at(16,2) {$\left\{
			\begin{array}{rcll}
				x_1+x_2+x_3  & \le & 2 \\
				x_1+x_2 + x_4 &\le & 2\\
				x_1 + x_3 + x_4 &\le &2 \\
				2x_1+x_2+x_3+x_4 & \le & 3 \\
				-x_i  &  \le &  0 & \forall i \in \{1,2,3,4\}
			\end{array}
			\right\}$};
		\draw (1) -- (2);
		\draw (1) -- (3);
		\draw (1) -- (4); 
	\end{tikzpicture}
	\caption{A star and the linear description of its co-2-plex polytope.}
	\label{fig:starg}
\end{figure}

\begin{corollary}\label{cor:tighterE}
	The ILP  $\mathcal{E}(G)$ has a tighter linear relaxation value than $\mathcal{N}_2(G)$. 
\end{corollary}
\noindent\begin{proof}
	Lemma~\ref{the: firstTreecharac} and the fact that the non trivial inequalities of $\mathcal{N}_2(G)$ belong to Inequalities~\eqref{iq:tree} yield that $\mathcal{E}(G)$ is tighter than $\mathcal{N}_2(G)$.
	We provide an example of a graph and a fractional point that is feasible for the linear relaxation of $\mathcal{N}_2(G)$ and infeasible for the linear relaxation of $\mathcal{E}(G)$.
	Consider the star with vertex $1$ as center and vertices $2$, $3$, $4$ as leaves, see Figure~\ref{fig:starg}. The point $\bar{x}_1 = \frac{1}{2}$, $\bar{x}_2 = \bar{x}_3 = 1$ and $\bar{x_4} = 0$ is a solution of the linear relaxation of $\mathcal{N}_2(G)$. However, $\overline{x}$ is not contained in $\mathcal{T}(G)$ since it violates the constraint \eqref{iq:tree} associated with node $1$ and neighbors $2$ and~$3$. By Lemma~\ref{the: firstTreecharac}, there exists no $y$ such that $(\bar{x},y)$ is a solution to the linear relaxation of $\mathcal{E}(G)$.
\end{proof}

Balasundaram et  al.~\cite {bala} propose to reinforce $\mathcal{N}_2(G)$ with Inequalities~\eqref{eq:2plex}. These latter cut off the fractional point provided in the previous proof. However, they are not sufficient to describe the co-2-plex polytope on stars.

\begin{theorem}\label{the:formutterclique}
	The ILP $\mathcal{K}(G)= \max \{w^\top x : (x,y) \in \{0,1 \}^{V+ E} \text{ satisfies \eqref{iq:cliquestorng}--\eqref{eq:triviale}} \}$ has a tighter linear relaxation value than $\mathcal{N}_2(G)$ reinforced with Inequalities~\eqref{eq:2plex} for finding maximum $w$-weighted co-2-plexes. 
\end{theorem}
\noindent\begin{proof}
	By Corollary~\ref{cor:tighterE} and Observation~\ref{obs:dominatedUtterCliques}, the linear constraint set of $\mathcal{K}(G)$ defines a polytope included in the polytope defined by the linear constraint set of $\mathcal{N}_2(G)$. It is now sufficient to show that Inequalities~\eqref{eq:2plex} are redundant with respect to \eqref{iq:cliquestorng}--\eqref{eq:triviale}. 
	
	 Let $K$ be a 2-plex of~$G$. By definition of 2-plexes, $K$ induces a graph built from a complete graph in which a matching has been removed. Taking one extremity of each edge of this matching yields a node set $K_1$. Then, $K_1$ is a clique of $G$, as well as $K_2 = K \setminus K_1$. Moreover, for all $v \in K_1$ and $e \in E(K_2)$, $v$ and $e$ are adjacent by contraction so $[K_1,E(K_2)]$ is an utter clique of $G$, and so is $[K_2,E(K_1)]$. Thus, in the sum of the Inequalities~\eqref{iq:cliquestorng} associated with $[K_1,E(K_2)]$ and $[K_2,E(K_1)]$, the $y$ variables vanish, and we obtain the inequality~\eqref{eq:2plex} associated with $K$.
	
	It remains to exhibit a graph for which the inclusion is strict. Consider the star in Figure~\ref{fig:starg}. The point $\bar{x}_i = \frac{2}{3}$ for $i = 1, \dots, 4$ belongs to the linear relaxation of $\mathcal{N}_2(G)$ reinforced by 2-plex inequalities. Indeed, each maximal 2-plex contains node $1$ and two nodes among $2$, $3$, and $4$, so all 2-plex inequalities are satisfied by $\bar{x}$.
	However, $\bar{x}$ violates the inequality \eqref{iq:tree} associated with node $w = 1$ and neighbor subset $W =\{ 2, 3,4 \}$. Lemma~\ref{the: firstTreecharac} proves that there exists no $y$ such that $(\bar{x},y)$ is a solution of the polytope defined by \eqref{iq:cliquestorng}--\eqref{eq:triviale}.
\end{proof}

\subsection{Exact separation for star Inequalities~\eqref{iq:tree} \label{sec:exactSeparationStarIneq}}

As a consequence of Observation~\ref{obs:iqedge} and Lemma~\ref{the: firstTreecharac}, we get that Inequalities~\eqref{iq:tree} are valid for formulation $\mathcal{N}_2(G)$.

\medskip

\begin{theorem}\label{the:separationstar}
Star Inequalities~\eqref{iq:tree} can be separated in polynomial time.
\end{theorem}
\noindent\begin{proof}	
	If a most violated constraint associated with vertex $w$ and its subset of neighbors $W$ is violated, then $x^*_w + x^*_u >1$ for all $u \in W$, as otherwise, for some vertex $u$, the inequality $(|W|-2)x_w + x(W\setminus \{u\}) \le |W|-1$ would yield a larger violation. 
	
	Moreover, if there exists a vertex $v \in N(w) \setminus W$ such that $x^*_v + x^*_u >1$ for all $u \in W$, then the inequality associated with $W \cup \{ v\}$ yields a larger violation than the one associated with $W$. For these reasons, $W = \{v \in N(w) : x_w^* + x_u^* > 1\}$ yields a most violated Inequality~\eqref{iq:tree} for a given $w$. Computing $W$ for each $w \in V$ is done in polynomial time. It can be checked as a post-processing step whether the violation is strictly positive. 
\end{proof}

A preliminary computation study showed that reinforcing $\mathcal{N}_2(G)$ with star inequalities using the upward separation yields a very bad algorithm in terms of computation time (reinforcing it with both 2-plexes and star inequalities is also very bad).

\section{Computational experiments}

In this section, we describe computational experiments we conduct to test the performance of the different formulations and inequalities we derive from Sections~\ref{sec:polyhedralContributions} and~\ref{sec:LPformulations}. We first describe the instance datasets and the different algorithms tested on these datasets. We then make an experimental comparison of the different formulations.

\subsection{Instance datasets}

We run experiments on the maximum cardinality variant of the co-2-plex problem. Hence, each instance is only described by a graph. We consider 212 instances divided into six datasets as follows.

\subparagraph{Dataset \texttt{Color02}:} the \texttt{Color02} dataset~\cite{color02} is a graph coloring dataset containing 74 graphs for which an optimal coloring can be computed in less than an hour.

\subparagraph{Datasets \texttt{DIMACS} and \textoverline{\texttt{DIMACS}}:} 
the dataset \texttt{DIMACS} contains the 9 instances from the 10$^{\rm th}$ DIMACS challenge~\cite{dimacs10} with less than 300 vertices. We also consider the dataset \textoverline{\texttt{DIMACS}} composed of the 9 instances obtained by complementing the graphs in \texttt{DIMACS}, since the co-2-plex has been solved on these instances in \cite{roberto2} when solving the pricing problem on a branch-and-price algorithm designed for the 2-defective coloring. 

\subparagraph{Random datasets \texttt{Random(0.3)}, \texttt{Random(0.5)} and \texttt{Random(0.7)}:} we consider 3 random datasets containing each 40 instances generated similarly to those in~\cite{HANSEN2009135}. More precisely, these instances are obtained by randomly generating graphs with 70, 80, 85, and 90 vertices (10 random graphs for each given number of vertices). Graphs in datasets \texttt{Random(0.3)}, \texttt{Random(0.5)} and \texttt{Random(0.7)} are generated following Erdős–Rényi models with a uniform probability to have an edge of 0.3, 0.5, and 0.7, respectively.

\medskip

Table~\ref{table:datasetStatistics} presents average statistics of the instances for each dataset. The average number of vertices is similar for all datasets except \texttt{Color02}, for which it is more than twice as high. The average density varies significantly between datasets. We highlight this variation since our extended formulations consider additional variables associated with edges of the graph. Hence, density is an important feature to consider when analysing the performance of the algorithms. 


\begin{table}[h]
\centering
\begin{tabular}{|l|r|r|r|r|}
\hline
     Instances set& av. vertices & av. density & av. min degree & av. max degree  \\
     \hline
     \texttt{Color02}& 221 &0.2& 15 & 90 \\
     \texttt{Random(0.3)} & 82 & 0.3 & 14&33 \\
     \texttt{Random(0.5)} & 82 & 0.5 & 29&51 \\
     \texttt{Random(0.7)} & 82 & 0.7 & 46&65 \\
     \texttt{DIMACS} & 93 & 0.9 & 52 & 89 \\
     \textoverline{\texttt{DIMACS}} & 93 & 0.1 & 2 & 40 \\
     \hline
     All &131&0.4 & 24 & 65 \\
     \hline
\end{tabular}
\caption{Statistics of the different instance sets.}\label{table:datasetStatistics}
\end{table}



\subsection{Compared algorithms}\label{sec:comparedalg}

In this section, we describe the different algorithms that we run on the datasets. 
All have been implemented in the Julia programming language~\cite{bezanson2017julia} using JuMP~\cite{Lubin2023} with CPLEX 20.1 as ILP solver with a time limit of one hour. 

Since experiments are conducted on the maximum cardinality co-2-plex problem, we use the preprocessing devised by Gao et. al in~\cite{doi:10.1142/S0218213019500155}. We first compute a co-2-plex $S$ by running the GRASP algorithm defined in~\cite{grasp}\footnote{The difference with the algorithm of \cite{grasp} is that we only perform one execution of the algorithm with $\alpha = 0.7$.}.
Then, starting with $W= V$, the preprocessing iteratively removes from $W$ vertices with a degree greater than or equal to $|W| - |S| + 1$ since these vertices cannot be in a co-2-plex of cardinality greater than the one of $S$. FInally, the problem is decomposed into $|W|$ smaller problems. For each $v \in W$, the smaller problem reduces to computing a maximum cardinality co-2plex containing $v$ in $G[W \cap \bar D^2_v]$, where $\bar D^2_v$ corresponds to the set of vertices that are at distance at most two of $v$ in $\bar G$, that is, there exists in $\bar G$ a path of length two between $v$ and each vertex of $\bar D^2_v$ non-adjacent to $v$. This preprocessing stems from the fact that, in the graph induced by a 2-plex of size at least three, vertices are at distance at most two. The measures reported in the experimental tables for each algorithm over an instance are the sum of the measures of this algorithm over the $|W|$ smaller instances obtained by this preprocessing.



\subparagraph{$\mathcal{N}_2(G)$ algorithm.} This algorithm consists in solving Formulation  $\mathcal{N}_2(G)$ using CPLEX with default parameters.

\subparagraph{$\mathcal{N}_2(G)$ + 2-plexes.} This consists in solving  Formulation  $\mathcal{N}_2(G)$  strengthened by 2-plex inequalities~\eqref{eq:2plex} using the algorithm of Balasundaram~\cite{bala} as described in Section~\ref{sec:LitteratureReviewILP}.

 \subparagraph{$\mathcal{E}(G)$ algorithm.} This algorithm consists of solving $\mathcal{E}(G)$  using CPLEX with default parameters.

 \subparagraph{$\mathcal{E}(G)$ + utter cliques.} This algorithm consists in solving $\mathcal{E}(G)$ reinforced with a heuristic separation of constraints~\eqref{iq:cliquestorng}.
 The separation is done by mapping the incumbent solution $(x^*,y^*)$ into a fractional solution $(z^*,y^*)$ of \eqref{ineq:tdi1}--\eqref{ineq:tdi3} using $\phi^{-1}$ and then computing a maximum clique using the same clique greedy algorithm as the one described in the separation phase of the algorithm of Balasundaram~\cite{bala}, see Section~\ref{sec:LitteratureReviewILP}.

\subsection{Computational results}

%
%
%
%
	

In Table~\ref{tabl:performancesco2plex} we give the results of the four algorithms respectively on each instance set and on all at once. Entries in the table are the following:

\begin{tabular}{ll}
    -- $\%$ solved & pourcentage of instance solved,\\
	-- CPU & average running time in seconds,\\
	-- \# nodes & average number of generated branching nodes,\\
	-- \# cuts & average number of generated reinforcement cuts,\\
    -- gap & average gap over the instance set where the
    gap for a single instance is the maximum\\
    & of the $|W|$ gaps computed for each element of the decomposition described in Section~\ref{sec:comparedalg}. \\
    & When the gap is closed for every instance of the set, we replace the value by '-'.\\
    -- root Gap & average gap at the root node.
\end{tabular}

\medskip

 $\mathcal{N}_2(G)$ is the fastest algorithm over all instance sets but  \textoverline{\texttt{DIMACS}}, where $\mathcal{E}(G)$ is the fastest one. These two algorithms respectively solve 190 and 189 instances over the 212, and clearly stand out from their reinforced version both in the number of instances solved and also in the CPU time. Unless for \texttt{Color02} where  $\mathcal{N}_2(G)$ + 2-plexes achieves the lowest gap, it is always either $\mathcal{N}_2(G)$ or $\mathcal{E}(G)$ that achieve the lowest gap. When considering the root node gap, we see that adding cuts does not necessarily reduce the gap. In fact, $\mathcal{N}_2(G)$ + 2-plex achieves a higher gap then $\mathcal{N}_2(G)$ does on \texttt{Color02}, \texttt{Random(0.3)} and \textoverline{\texttt{DIMACS}}. Moreover, $\mathcal{E}(G)$ + utter cliques always achieves a higher gap than $\mathcal{E}(G)$ does. This phenomenon shows that adding reinforcement cuts perturbs CPLEX's generic reinforcement procedures.  However, even if $\mathcal{E}(G)$ + utter cliques does not perform well on most instance sets, it solves two more instances than the other algorithms on \texttt{Color02}.

One can observe that the lower the density of the instances set is, the better the performances of $\mathcal{E}(G)$ and $\mathcal{E}(G)$ + utter cliques are. Indeed, $\mathcal{E}(G)$ is the fastest algorithm over \textoverline{\texttt{DIMACS}}, which is the sparsest set of instances. This mainly has two explanations. First, the number of variables of $\mathcal{E}(G)$ is larger for denser instances. Moreover, Theorem~\eqref{the:facettes} suggests that star inequalities are facet-defining only when the neighborhood of the vertex considered is sparse, and by the projection of Theorem~\ref{the:contractionperfectIntegralPol}'s proof, we know that Inequalities~\eqref{iq:edgesExtended} can be facet-defining only if star inequalities are. This lets us think that the constraints set of $\mathcal{E}(G)$ should also be tighter when the graph is sparse. The root gaps reflect this observation as the only instances set where $\mathcal{N}_2(G)$ achieves a lower gap than $\mathcal{E}(G)$, \texttt{Random(0.7)} is also the densest set of instances. The latter is also the single set of instances on which reinforcing $\mathcal{E}$ with several utter clique constraints does not reduce the number of branching nodes. 

Overall, even if adding cuts does not reduce the root node gap of a formulation, the theoretical comparison of the linear relaxations of $\mathcal{E}(G)$ and $\mathcal{N}_2(G)$ is confirmed by at least two indicators: the root gap and the number of branching nodes. The theoretical comparison of $\mathcal{E}(G)$ + utter cliques and $\mathcal{N}_2(G)$ + 2-plex is only confirmed by the reduction of the number of branching nodes, the root gap of $\mathcal{E}(G)$ + utter cliques is only lower on the sparser sets of instances: \texttt{Color02}, \texttt{Random(0.3)} and \textoverline{\texttt{DIMACS}}.

\medskip

\begin{table}[h!] 
 \centering 
 \begin{tabular}{|l|l|r|r|r|r|r|r|} 
 \hline 
 Dataset & Algorithm  &\% solved & CPU  & \# nodes & \# cuts & gap & root Gap\\ 
 \hline
\multirow{6}{*}{\texttt{Color02}}
& $\mathcal{N}_2(G)$&54/74&1048&7.9e6&-&0.15&0.75\\
&$\mathcal{E}(G)$ &54/74&1100&3.8e5&-&0.14&0.36\\
& $\mathcal{N}_2(G)$ + 2-plexes &53/74&1137&1.1e6&5.0e4&0.06&1.29\\
&$\mathcal{E}(G)$ + utter cliques  &56/74&1120&9.1e3&5.2e3&0.20&0.51\\
\hline
\multirow{6}{*}{\texttt{Random(0.3)}}
&$\mathcal{N}_2(G)$&40/40 &878&1.6e7&-&-&1.17\\
& $\mathcal{E}(G)$ &40/40&1207&7.2e5&-&-&0.61\\
&$\mathcal{N}_2(G)$ + 2-plexes &16/40&2605&7.3e6&1.8e5&0.20&1.62\\
&$\mathcal{E}(G)$ + utter cliques &31/40&1972&1.3e5&7.3e4&0.12&1.20\\
\hline
\multirow{6}{*}{\texttt{Random(0.5)}}
&$\mathcal{N}_2(G)$ &40/40 &145&3.1e6&-&-&1.92\\
&$\mathcal{E}(G)$ &40/40&573&2.0e5&-&-&1.55\\
&$\mathcal{N}_2(G)$ + 2-plexes &40/40&603&1.6e6&8.1e4&-&1.87\\
& $\mathcal{E}(G)$ + utter cliques &22/40&2532&1.0e5&6.1e4& 0.62&2.78\\
\hline
\multirow{6}{*}{\texttt{Random(0.7)}}
&$\mathcal{N}_2(G)$&40/40 &29&3.6e5&-&-&1.62\\
&$\mathcal{E}(G)$ &40/40&502&5.2e4&-&-&2.08\\
&$\mathcal{N}_2(G)$ + 2-plexes &40/40&251&3.0e5&6.0e5&-&1.24\\
& $\mathcal{E}(G)$ + utter cliques &20/40&2866&6.5e4&3.7e4&0.89&3.18\\
\hline
\multirow{6}{*}{\texttt{DIMACS}}
&$\mathcal{N}_2(G)$& 9/9&197&5.9e5&-&-&1.58\\
&$\mathcal{E}(G)$ &7/9&821&5.8e3&-&1.06&0.20\\
&$\mathcal{N}_2(G)$ + 2-plexes &9/9&322&1.5e5&3.2e4&-&0.25\\
& $\mathcal{E}(G)$ + utter cliques &7/9&992&1.8e3&2.9e3&0.70&0.99\\
\hline
\multirow{6}{*}{\textoverline{\texttt{DIMACS}}}
&$\mathcal{N}_2(G)$&7/9 &1020&4.9e6&-&0.02&0.18\\
&$\mathcal{E}(G)$ &8/9&916&9.2e4&-&0.01&0.06\\
&$\mathcal{N}_2(G)$ + 2-plexes &6/9&1214&9.1e5&6.6e4&0.03&0.77\\
& $\mathcal{E}(G)$ + utter cliques &7/9&935&2.5e4&7.0e3&0.01&0.15\\
\hline 
\multirow{6}{*}{\texttt{All instances}}
&$\mathcal{N}_2(G)$&190/212 &616&6.8e6&-&0.05&1.23\\
&$\mathcal{E}(G)$ & 189/212 &888&3.2e5&-&0.10&0.94\\
&$\mathcal{N}_2(G)$ + 2-plexes &164/212 &1115&2.2e6&8.4e4&0.06&1.38\\
& $\mathcal{E}(G)$ + utter cliques & 143/212 &1863&6.1e4&3.4e4&0.40&1.58\\
\hline 
 \end{tabular} 

 \caption{Relative performances of the algorithms.}
	\label{tabl:performancesco2plex}
 \end{table}

\medskip

 For clarity's sake, we omit the details of preliminary experiments corresponding to reinforcing $\mathcal{N}_2(G)$ and $\mathcal{N}_2(G)$ + 2-plex by star inequalities using the separation algorithm described in the proof of Theorem~\ref{the:separationstar}. The two corresponding algorithms yielded poor performances. Similarly, we tried to solve $\phi^{-1} (\mathcal{E}(G))$ and $\phi^{-1} (\mathcal{E}(G) + \text{ utter cliques })$ and both algorithms yielded poorer performances than $\mathcal{E}(G)$ and $\mathcal{E}(G)$ + utter cliques.

\section*{Conclusion}

In this article, we have shown how extended formulations for the maximum weighted co-2-plex polytope of a graph naturally arise from the stable set polytope of its utter graph. This permits us to give new characterizations of contraction perfect graphs based on TDI systems and integer polytopes. We projected an extended space description for the co-2-plex polytope of trees, which exhibited a new family of facet-defining inequalities for the co-2-plex polytope of any graph. We characterized when they suffice to describe the polytope, which turned out to hold for trees but also for holes of length multiple of 3. As these inequalities come in exponential numbers, we proposed
a polynomial separation.
 From these polyhedral results, we derived new compact ILPs for the maximum co-2-plex problem, which we tried to reinforce with a greedy separation algorithm. These new formulations allowed us to make a computational comparison, showing that the extended formulations do not yield a faster algorithm even if they yield a much smaller number of branching nodes, confirming the theoretical comparison of their respective linear relaxation. The extended formulation solves sparse instances more efficiently, and overall $\mathcal{N}_2(G)$ still yields the faster algorithm.

\bibliographystyle{acm}
\bibliography{bibtex.bib}

\end{document}